\documentstyle[12pt]{article}
\date{}
\def\be{\begin{equation}}
\def\ee{\end{equation}}
\def\bea{\begin{eqnarray}}
\def\eea{\end{eqnarray}}
\def\s{\sigma}
\def\al{\alpha}
\def\lm{\lambda}
\def\de{\delta}
\def\om{\omega}
\def\pr{\prime}
\title{Initial-boundary value problem and stability of
solutions for string baryon model ``triangle"}
\author{G.\,S. Sharov\thanks{E-mail: german.sharov@tversu.ru}, V.\,P. Petrov\\
{\small Tver state university}\\
{\small Tver, 170002, Sadovyj per. 35, Mathem. dep-t.}}
\begin{document}
\maketitle
\begin{abstract}

For the string baryon model ``triangle" the initial-boundary value problem
is stated and solved in general.
This problem implies defining a classical motion of the system on the base
of given initial position and initial velocities of string points.
The presented solution reduces the initial-boundary value problem
for the considered model to the system of ordinary differential
equations that can be integrated numerically in general.
Using this approach we ascertain the stability of the
rotational motions (flat uniform rotations) for the ``triangle"
string configuration.
\end{abstract}

\section*{Introduction}

For describing orbitally excited baryons various string models are used.
They differ from each other in the topology of spatial junction
of three massive points (quarks) by relativistic strings.
Four variants of this junction are possible (Fig.\,1):
(a) the quark-diquark model q-qq \cite{Ko} (from the point of view
of classical dynamics it coincides with the meson model of relativistic
string with massive ends \cite{Ch,BN});
(b) the linear configuration q-q-q \cite{4B,lin};
(c) the ``three-string" model or Y-configuration \cite{AY,PY}
and (d) the ``triangle" model or $\Delta$-configuration \cite{Tr,PRTr}
that is under consideration in this paper.

\begin{figure}[h]
\begin{center}
\begin{picture}(150,28)
\thicklines
\put(10,15){\line(1,0){25}}
\put(10,15){\circle*{2}}\put(35,16){\circle*{2}}\put(35,14){\circle*{2}}
\put(9,11){q}\put(32,9){qq}\put(12,26){(a)}
\put(50,15){\line(1,0){28}}
\put(50,15){\circle*{2}}\put(64,15){\circle*{2}}\put(78,15){\circle*{2}}
\put(49,11){q}\put(63,11){q}\put(77,11){q}\put(52,26){(b)}
\put(98,15){\line(0,-1){12}}\put(98,15){\line(-2,1){10}}
\put(98,15){\line(2,1){10}}
\put(98,3){\circle*{2}}\put(88,20){\circle*{2}}\put(108,20){\circle*{2}}
\put(94,3){q}\put(85,17){q}\put(109,17){q}\put(93,26){(c)}
\put(115,5){\line(1,0){24}}\put(115,5){\line(2,3){12}}
\put(127,23){\line(2,-3){12}}
\put(115,5){\circle*{2}}\put(139,5){\circle*{2}}\put(127,23){\circle*{2}}
\put(110,5){q}\put(141,5){q}\put(123,25){q}\put(132,26){(d)}
\end{picture}\end{center}
\caption{String baryon models.}
\end{figure}
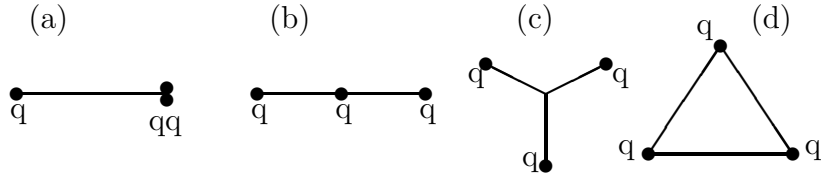

The exact solutions of the classical equations of motion describing
uniform rotations of the system are known for all these models.
For the configurations q-qq and q-q-q the rotating string has the form
of a rectilinear segment \cite{Ko,Ch,BN}.
For the ``three-string" model there are three rectilinear segments
joined in a plane at the angles 120${}^\circ$ \cite{AY,PY}.
For the model ``triangle" the rotating string has the form of the closed
curve consisting of segments of a hypocycloid \cite{Tr,PRTr}.

The connection between the energy of the system $E=M$ and its angular
momentum $J$ for all these motions in the high energy limit
(for any way of taking quark spins into account or neglecting quark spins)
has the form \cite{Ko,4B,PRTr,Solov}
\be
J\simeq\al'E^2,\qquad J,E\to\infty.\label{je}\ee
This fact allows us to apply each of the mentioned models to
describing the baryon states with large $J$ on the Regge
trajectories \cite{4B}.

The problem of choosing the most adequate model among the four mentioned ones
is not solved yet. For the q-qq and q-q-q configurations the Regge
slope $\al'$ in Eq.~(\ref{je}) is connected with the string tension
$\gamma$ by the Namby relation $\al'=(2\pi\gamma)^{-1}$ \cite{Nambu}
and the tension should be equal for mesons and baryons. The configurations
Y and $\Delta$ describe the Regge trajectories if the effective tension
$\gamma$ for them is lower then the ``mesonic" tension
$\gamma\simeq0.18\mbox{ GeV}^2$ \cite{4B}.

On the other hand the QCD-motivated baryon Wilson loop operator approach
gives some arguments in favour of the Y-configuration \cite{Isg} or
the ``triangle" model \cite{Corn}.

When we choose the adequate string baryon configuration we are also take into
account the stability of rotational motions for these systems.
In particular, the rotational motions of the q-q-q system with the middle
quark at rest are unstable with respect to centrifugal moving away
of the middle quark \cite{lin}. Any small disturbance results in
a complicated quasiperiodic motion, but the system doesn't transform
into the quark-diquark one \cite{Ko}.
The latter q-qq configuration (or the meson string model)
seems to be stable but this question is not exhaustively studied.
Small disturbances of the well known exact solution (uniformly rotating
rectilinear string) result in a rotation of the string with slightly
varying shape and interquark distance \cite{JVM}. Some small corrections
to the meson string rotational motions are searched in
Refs.~\cite{AllenOlssV}.

The stability problem for the other string baryon models (Fig.\,1)
wasn't studied. In this paper we investigate in this respect the
hypocycloidal rotational motions in the model ``triangle" \cite{Tr,PRTr}.

There is the opinion that an orbitally excited baryon system
with the given momentum $J$ ``chooses" the only steady state
with the lowest energy \cite{Ko}. In according with this statement
the simplest nontrivial rotational motion in the model ``triangle"
(three quarks connected by smooth segments of hypocycloids) should
transform into the ``quark-diquark" state \cite{PRTr} with two
quarks merging into one.

The simplest way to solve the stability problem is to develop a method
of solution of the initial-boundary value problem with arbitrary
initial conditions for the ``triangle" model. Such a method is suggested
in this paper.

In Sect.\,1 the equations of motion with their common solution and
the boundary conditions for the string baryon model ``triangle" are
given.
In Sect.\,2 the solution of the initial-boundary value problem is suggested.
It develops the approach used earlier for the meson string model
with massive ends \cite{BaSh} and for the q-q-q baryon configuration \cite{lin}.
In the last section the stability of the rotational motions is verified.

\section{Dynamics of string baryon model ``triangle"}

Let us consider a closed relativistic string with the tension $\gamma$
carrying three pointlike masses $m_1$, $m_2$, $m_3$.
The action for this system is \cite{Tr,PRTr}
\be
S=-\gamma\int\limits_{\tau_1}^{\tau_2}\!d\tau\!\!
\int\limits_{\s_0(\tau)}^{\s_3(\tau)}\!\!\!\sqrt{-g}\,d\s-
\sum_{i=1}^3m_i\!\int\limits_{\tau_1}^{\tau_2}\!\left\{\big[{\textstyle
\frac d{d\tau}X^\mu\big(\tau,\s_i(\tau)\big)}\big]^2\right\}^{1/2}d\tau.
\label{S}\ee
Here $X^\mu(\tau,\s)$ are coordinates of a point of
the string in $D$-dimensional Minkowski space $R^{1,D-1}$ with
signature $+,-,-,\dots$, $g=\dot X^2X'{}^2-(\dot XX')^2$
$\,(\tau,\s)\in\Omega=\Omega_0\cup\Omega_1\cup\Omega_3$ (Fig.\,2),
$(a,b)=a^\mu b_\mu$ is the (pseudo)scalar product,
$\dot X^\mu=\partial_\tau X^\mu$, $X'{}^\mu=\partial_\s X^\mu$,
the speed of light  $c=1$; $\s_i(\tau)$ are
inner coordinates of the quark world lines\footnote{We use the term
``quark" instead of ``material point" for brevity, keeping in mind that
the considered model on the classic level doesn't describe the spin
and other quantum numbers of quarks.}, $i=0,\,1,\,2,\,3$.

The equations $\s=\s_0(\tau)$ and $\s=\s_3(\tau)$
determine the trajectory of the same (third) quark. It is connected with
the fact that string is closed and may be written in the following general
form \cite {Tr}:
\be
X^\mu(\tau,\s_0(\tau))=X^\mu(\tau^*,\s_3(\tau^*)).\label{cl}\ee

The parameters $\tau$ and $\tau^*$ in these two parametrizations
of just the same line are not equal in general.

Variation and minimization of action (\ref{S}) result in the
equations of motion of the string and the boundary conditions at the
quark trajectories. These trajectories are smooth curves.
Derivatives of $X^\mu$ can have discontinuities on the
lines $\s=\s_i(\tau)$ (except for tangential).

However, through a nondegenerate reparametrization
$\tau=\tau(\tilde\tau,\tilde\s)$, $\s=\s(\tilde\tau,\tilde\s)$
the induced metric on the world surface of the string may be made
continuous and conformally flat \cite {Tr}, i.e.,
satisfying the orthonormality conditions
\be
\dot X^2+X'{}^2=0,\qquad(\dot X,X')=0.\label{ort}\ee
Under conditions (\ref{ort}) the equations of motion become linear
\be
\ddot X^\mu+X''{}^\mu=0,\label{eq}\ee
and the boundary conditions take the simplest form \cite{Tr,PRTr}
\be
m_i\frac d{d\tau}U_i^\mu(\tau)
-\gamma\big[X'{}^\mu+\s_i'(\tau)\,\dot X^\mu\big]
\Big|_{\s=\s_i+0}\!\!
+\gamma\big[X'{}^\mu+\s_i'(\tau)\,\dot X^\mu\big]
\Big|_{\s=\s_i-0}\!\!=0.
\label{qqq}\ee
Here the notation
\be
U^\mu_i(\tau)=\frac{\dot X^\mu+\s_i'(\tau)\,X'{}^\mu}
{\sqrt{\dot X^2\cdot(1-\s_i'{}^2)}}\bigg|_{\s=\s_i(\tau)}
\label{U}\ee
for unit $R^{1,D-1}$-velocity vector of i-th quark is used.

In accordance with the closure condition (\ref{cl}) for the third quark $i=3$
one should put $\s=\s_0(\tau)$ in the second summand in Eq.~(\ref{qqq})
and replace $\tau$ by $\tau^*$ in the last term.

At this stage we have five undetermined functions in this model:
$\tau^*(\tau)$ in the closure condition (\ref{cl}) and four
trajectories $\s_i(\tau)$, $i=0,\,1,\,2,\,3$.
Using the fact that Eqs. (\ref{ort})\,--\,(\ref{qqq})
are invariant with respect to the reparametrizations \cite{BN}
\be
\tau\pm\s=f_\pm(\tilde\tau\pm\tilde\s).\label{rep}\ee
we fix two\footnote{The remaining three functions will be calculated
with solving the initial-boundary value problem in Sects.\,2, 3.}
of the mentioned functions as follows (Fig.\,2):
\be\s_0(\tau)=0,\qquad\tau^*(\tau)=\tau.
\label{fit}\ee
The first Eq. (\ref{fit}) may be obtained at the first step
through the above reparametrization (\ref{rep}) with the required $f_+$
and $f_-(\eta)= \eta$. At the second step one can get the equality
$\tau^*=\tau$ by repeating the procedure (\ref{rep}) with the function
$f_+=f_-\equiv f$ that satisfies the condition
$$2f(\tau)=f\big(\tau^*(\tau)+\s_3(\tau)\big)+
f\big(\tau^*(\tau)-\s_3(\tau)\big).$$
The equality $\tau^*(\tau)=\tau$ is equivalent to the closure of any coordinate
line $\tau={}$const on the world surface.

\begin{figure}[hb]
\begin{center}
\begin{picture}(160,60)
\put(15,2){\vector(1,0){140}} \put(20,0){\vector(0,1){59}}
\put(17,3){0} \put(154,4){$\s$} \put(21,59){$\tau$}
\put(6,20){$\s_0=0$} \put(65,28){$\s_1(\tau)$}
\put(112,50){$\s_2(\tau)$} \put(144,25){$\s_3(\tau)$}
\put(41,37){$\Omega_0$}
\put(84,37){$\Omega_1$} \put(125,37){$\Omega_2$}
\put(40,14){$A_0$}\put(83,15){$A_1$}\put(115,15){$A_2$}
\put(20,2){\line(1,1){45.5}} \put(138,2){\line(-1,1){31.5}}
\put(63,5){\line(1,1){48}} \put(63,5){\line(-1,1){43}}
\put(102,7){\line(-1,1){36}} \put(102,7){\line(1,1){47.5}}
\thicklines
\put(20,2){\line(0,1){51}}
\put(20,2){\line(5,-1){5}} \put(25,1){\line(6,-1){6}}
\put(31,0){\line(1,0){4}} \put(35,0){\line(6,1){6}}
\put(41,1){\line(5,1){10}} \put(51,3){\line(6,1){24}}
\put(75,7){\line(5,1){5}} \put(80,8){\line(6,1){6}}
\put(86,9){\line(1,0){5}} \put(91,9){\line(6,-1){6}}
\put(97,8){\line(5,-1){10}} \put(107,6){\line(6,-1){12}}
\put(119,4){\line(1,0){7}} \put(126,4){\line(6,-1){12}}
\put(63,5){\line(-1,6){1}} \put(62,11){\line(0,1){6}}
\put(62,17){\line(1,6){2}} \put(64,29){\line(1,5){1}}
\put(65,34){\line(1,6){1}} \put(66,40){\line(0,1){5}}
\put(66,45){\line(-1,6){2}}
\put(102,7){\line(0,1){4}} \put(102,11){\line(1,6){2}}
\put(104,23){\line(1,5){1}} \put(105,28){\line(1,4){1.5}}
\put(106,32){\line(1,3){2}} \put(108,38){\line(1,4){2}}
\put(110,46){\line(1,5){1}} \put(111,51){\line(1,6){1}}
\put(138,2){\line(1,6){2}} \put(140,14){\line(1,5){1}}
\put(141,19){\line(1,4){2}} \put(143,27){\line(1,3){2}}
\put(145,33){\line(1,4){2}} \put(147,41){\line(1,5){1}}
\put(148,46){\line(1,6){2}}
\end{picture}
\caption{The domain $\Omega$ and the initial curve.}
\end{center}
\end{figure}
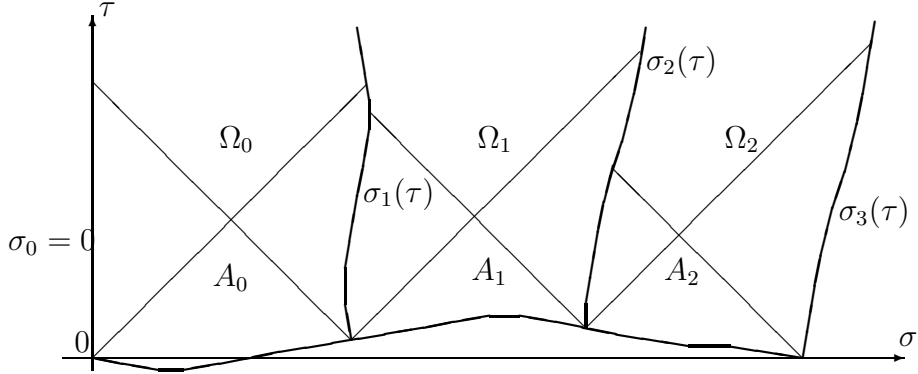

The domain $\Omega$ is divided into three ones
$\Omega_i=\{(\tau,\s):\,\s_i(\tau)<\s<\s_{i+1}(\tau)\}$, $i=0,\,1,\,2$
(Fig.\,2) by the quark trajectories. The general solutions of Eq.~(\ref{eq})
in these domains are described by different functions
\be
X^\mu(\tau,\s)=\frac1{2}\bigl[\Psi^\mu_{i+}(\tau+\s)+\Psi^\mu_{i-}(\tau-\s)
\bigr],\qquad (\tau,\s)\in\Omega_i,\qquad i=0,\,1,\,2.
\label{sol}\ee
It is the consequence of discontinuities of $X^{\pr\mu}$ at $\s=\s_i(\tau)$.
Nevertheless, the function $X^\mu$ and its tangential derivative
$\frac d{d\tau}X^\mu\big(\tau,\s_i(\tau)\big)$ are continuous on the
lines $\s_i(\tau)$. It results in the relations
\bea
&\Psi^\mu_{(i-1)+}(+)+\Psi^\mu_{(i-1)-}(-)=
\Psi^\mu_{i+}(+)+\Psi^\mu_{i-}(-),&\label{con1}\\
&\frac d{d\tau}\big[\Psi^\mu_{(i-1)+}(+)+\Psi^\mu_{(i-1)-}(-)\big]=
\frac d{d\tau}\big[\Psi^\mu_{i+}(+)+\Psi^\mu_{i-}(-)\big].&
\label{con}\eea
We use here and below the notations
$(+)\equiv\big(\tau+\s_i(\tau)\big)$, $(-)\equiv\big(\tau-\s_i(\tau)\big)$.

For the third quark ($i=3$) the equality (\ref{con1}) is
the closure condition (\ref{cl}).
Under conditions (\ref{fit}) this equality takes the form
$$\Psi^\mu_{2+}\big(\tau+\s_3(\tau)\big)+
\Psi^\mu_{2-}\big(\tau-\s_3(\tau)\big)=
\Psi^\mu_{0+}(\tau)+\Psi^\mu_{0-}(\tau).$$

Similarly, for $i=3$ in the r.h.s. of (\ref{con}) and anywhere below
one should replace the expressions $(1\pm\s_3)\Psi^{\pr\mu}_{3\pm}(\pm)$
by $\Psi^{\pr\mu}_{0\pm}(\tau)$ in accordance with
the conditions (\ref{cl}) and (\ref{fit}).

If we substitute the general solution (\ref{sol}) into the boundary
conditions (\ref{qqq}) and Eq. (\ref{U}) these equations take the form
\bea
&\begin{array}{c}\displaystyle
\frac{dU^\mu_i(\tau)}{d\tau}+\frac{\gamma}{2m_i}\bigg[(1+\s'_i)\big[
\Psi^{\pr\mu}_{(i-1)+}(+)-\Psi^{\pr\mu}_{i+}(+)\big]+{}\\
\displaystyle{}+(1-\s'_i)\big[\Psi^{\pr\mu}_{i-}(-)-
\Psi^{\pr\mu}_{(i-1)-}(-)\big]\bigg]=0,\end{array}&\label{qi}\\
&\begin{array}{c}\displaystyle
U^\mu_i(\tau)=\frac{(1+\s'_i)\,\Psi^{\pr\mu}_{(i-1)+}(+)+(1-\s'_i)\,
\Psi^{\pr\mu}_{(i-1)-}(-)}{\sqrt{2(1-\s^{\pr2}_i)\big(
\Psi^\pr_{(i-1)+}(+),\Psi^\pr_{(i-1)-}(-)\big)}}={}\\
\displaystyle{}=\frac{(1+\s'_i)\,\Psi^{\pr\mu}_{i+}(+)+
(1-\s'_i)\,\Psi^{\pr\mu}_{i-}(-)}{\sqrt{2(1-\s^{\pr2}_i)
\big(\Psi^\pr_{i+}(+),\Psi^\pr_{i-}(-)\big)}}.
\end{array}&
\label{Ui}\eea

\section{Initial-boundary value problem}

The initial-boundary value problem for the ``triangle" string
configuration is reduced to obtaining the solution
of Eq.~(\ref{eq}) sufficiently smooth \cite{lin,BaSh} in the domain
$\Omega=\bigcup\limits_i\Omega_i$ and satisfying the orthonormality
conditions (\ref{ort}), boundary conditions (\ref{qqq}) and
two given initial conditions: an initial position of the string
in Minkowski space and initial velocities of string points.

An initial position of the string can be given as a curve
in Minkowski space
\be x^\mu=\rho^\mu(\lm), \quad\lm\in [\lm_0,\lm_3].\label{rho}\ee
This curve is space-like ($[\rho^\pr(\lm)]^2<0$) and closed:
$\rho^\mu(\lm_0)=\rho^\mu(\lm_3)$. The function $\rho^\mu(\lm)$
is piecewise smooth, $\rho^{\pr\mu}$ may have discontinuities
at the quark positions $\lm=\lm_1,\,\lm_2$.

Initial velocities on the initial curve can be given as a time-like
vector $\,v^\mu(\lm)$, $\lm\in[\lm_0,\lm_3]$, $v^\mu(\lm)$ may be
multiplied by an arbitrary scalar function $\phi(\lm)>0$.
The condition $v^\mu(\lm_0)=v^\mu(\lm_3)\cdot{}$const is fulfilled.

To solve the problem we set parametrically the initial curve
(\ref{rho}) on the world surface (Fig.\,2)
$$\tau=\tau(\lm), \quad\s=\s(\lm), \qquad\lm\in[\lm_0,\lm_3]$$
an use the following general form for the initial position  and
velocities \cite{BaSh}:
\bea
&X^\mu\bigl(\tau(\lm),\s(\lm)\bigr)=\rho^\mu(\lm),&\label{X=r}\\
&\al(\lm)\,\dot X^\mu\bigl(\tau(\lm),\s(\lm)\bigr)+\beta(\lm)\,
X'^\mu\bigl(\tau(\lm),\s(\lm)\bigr)=v^\mu(\lm).\label{X=v}&\eea
Here $\al(\lm)$, $\beta(\lm)$ are arbitrary functions
satisfying the inequality $\al(\lm)>|\beta(\lm)|$.

Functions $\tau(\lm)$, $\s(\lm)$, $\al(\lm)$, $\beta(\lm)$ are related
by the formulas \cite{BaSh}
\be\s^\pr(\lm)=\frac{\al\Delta+\beta P}{v^2},\qquad
\tau^\pr(\lm)=\frac{\beta\Delta+\al P}{v^2},
\label{sita}\ee
where $\,P(\lm)=\big(v(\lm),\rho^\pr(\lm)\big)$,
$\,\Delta(\lm)=\sqrt{(v,\rho^\pr)^2-v^2\rho^{\pr2}}$.

The freedom in choosing the functions $\al(\lm)$, $\beta(\lm)$ or $\tau(\lm)$
and $\s(\lm)$ is connected with the existence of the class of the
reparametrizations (\ref{rep}) preserving the conditions (\ref{fit}) \cite{PeSh}.
When we choose the functions $\al(\lm)$, $\beta(\lm)$ we take into account
only the inequality $\al>|\beta|$ and conditions (\ref{fit}) resulting in
the constraints for $\beta(\lm)$ and $\tau(\lm)$
\be\beta(\lm_0)=0,\quad\s(\lm_0)=0,\quad\tau(\lm_0)=\tau(\lm_3)=0.
\label{bet}\ee

One may obtain the equalities $\tau(\lm_0)=\s(\lm_0)=0$ (Fig.\,2) by choosing
the constants of integration in Eq.~(\ref{sita}), and the equality $\tau(\lm_0)=\tau(\lm_3)$ in (\ref{bet}) --- through multiplying $\al(\lm)$ or $\beta(\lm)$ by a constant.

Using the formulas \cite{BaSh}
\be \Psi^{\pr\mu}_{i\pm}\bigl(\tau(\lm)\pm\s(\lm)\bigr)=
\frac{(\Delta\mp P)v^\mu\pm v^2\rho^{\pr\mu}}
{\Delta\bigl(\al(\lm)\pm\beta(\lm)\bigr)}
\label{psi}\ee
we can determine the functions $\Psi^{\pr\mu}_{i\pm}$
from the initial data in the following finite segments:
\be\begin{array}{l}
\Psi^\mu_{i+}(\xi),\quad\xi\in\bigl[\tau(\lm_i)+\s(\lm_i),
\tau(\lm_{i+1})+\s(\lm_{i+1})\bigr],\\
\Psi^\mu_{i-}(\xi),\quad\xi\in\bigl[\tau(\lm_{i+1})-\s(\lm_{i+1}),
\tau(\lm_i)-\s(\lm_i)\bigr],\rule[3mm]{0mm}{1mm}\end{array}
\label{pseg}\ee
that lets us find the solution of the problem in the form (\ref{sol})
in the zones (Fig.\,2)
$$A_i=\bigl\{(\tau,\s):\,\tau(\lm)<\tau\leq\s+\tau(\lm_i)-\s(\lm_i);\,
\tau+\s\leq\tau(\lm_{i+1})+\s(\lm_{i+1})\bigr\}.$$
In these zones the solution depends only on initial data
without influence of the boundaries.
The integration constants in Eq.~(\ref{psi}) are determined from the initial
condition (\ref{X=r}).
In others parts of the domains $\Omega_i$ the solution will be obtained
by the prolongation of the functions $\Psi^\mu_{i\pm}$ with the help of the
boundary conditions (\ref{qi}), (\ref{Ui}).

We transform the systems of the ordinary differential equations
(\ref{qi}), (\ref{Ui}) to the normal form by the method used in
the paper \cite{BaSh}.

Multiplying scalarly Eq.~(\ref{Ui}) by
$(1\pm\s'_i)\,\Psi^{\pr\mu}_{(i-1)\pm}(\pm)$,
$\ (1\pm\s'_i)\,\Psi^{\pr\mu}_{i\pm}(\pm)$ and taking into account the isotropy
of the vectors $\Psi^{\pr\mu}_{i\pm}$
$$\Psi^{\pr2}_{i+}(\tau)=\Psi^{\pr2}_{i-}(\tau)=0,\qquad i=0,1,2,$$
(resulting from conditions (\ref{ort})) we obtain the equalities

\be\begin{array}{c}
\frac1{\sqrt2}{\sqrt{(1-\s^{\pr2}_i)
\big(\Psi^\pr_{(i-1)+}(+),\Psi^\pr_{(i-1)-}(-)\big)}}=
(1+\s'_i)\big(U_i(\tau),\Psi'_{(i-1)+}(+)\big)=\\
=(1-\s'_i)\big(U_i(\tau),\Psi'_{(i-1)-}(-)\big)=
\frac1{\sqrt2}{\sqrt{(1-\s^{\pr2}_i)
\big(\Psi'_{i+}(+),\Psi'_{i-}(-)\big)}}=\\
=(1+\s'_i)\big(U_i(\tau),\Psi'_{i+}(+)\big)=
(1-\s'_i)\big(U_i(\tau),\Psi'_{i-}(-)\big).\rule[3mm]{0mm}{1mm}
\label{ch}\end{array}\ee
Eqs.~(\ref{Ui}) and (\ref{ch}) result in the relations
$$\!\!\!\!(1+\s'_i)\,\Psi^{\pr\mu}_{j+}(+)+(1-\s'_i)\,\Psi^{\pr\mu}_{j-}(-)=
2(1\pm\s'_i)\big(U_i(\tau),\Psi'_{k\pm}(\pm)\big)\,U_i^\mu(\tau)$$
taking place for all 8 variants of choosing $j$, $k$ equal to $i$ or $i-1$
and the sign $\pm$. The latter relations let us transform the summands
in (\ref{qi}) as follows:
$$\begin{array}{c}
\!(1+\s'_i)\,\Psi^{\pr\mu}_{(i-1)+}(+)
-(1-\s'_i)\,\Psi^{\pr\mu}_{(i-1)-}(-)=
2(1-\s'_i)\,\Psi^{\pr\nu}_{(i-1)-}(-)(U_i^\mu U_{i\nu}-\de^\mu_\nu),\\
(1-\s'_i)\,\Psi^{\pr\mu}_{i-}(-)-(1+\s'_i)\,\Psi^{\pr\mu}_{i+}(+)=
2(1+\s'_i)\,\Psi^{\pr\nu}_{i+}(+)(U_i^\mu U_{i\nu}-\de^\mu_\nu).
\rule[3mm]{0mm}{1mm}\end{array}$$
where $\de^\mu_\nu=\left\{\begin{array}{ll}1,&\mu=\nu\\0,&\mu\ne\nu.
\end{array}\right.$
Eqs.~(\ref{qi}) take the normal form now
\be\frac{dU^\mu_i}{d\tau}=\frac{\gamma}{m_i}\big[\de^\mu_\nu-U^\mu_i(\tau)\,
U_{i\nu}(\tau)\big]\big[(1+\s'_i)\,\Psi^{\pr\nu}_{i+}(+)+
(1-\s'_i)\,\Psi^{\pr\nu}_{(i-1)-}(-)\big].
\label{Uqi}\ee
As was mentioned above for $i=3$ in Eq.~(\ref{Uqi}) one should replace
$(1+\s'_3)\Psi^\mu_{3+}(+)$ by $\Psi^\mu_{0+}(\tau)$ in accordance with
Eqs.~(\ref{cl}) and (\ref{fit}).
The systems (\ref{Uqi}) need the initial conditions
\be  U^\mu_i\big(\tau(\lm_i)\big)=v^\mu(\lm_i)\big/\sqrt{v^2(\lm_i)},
\qquad i=1,2,3.\label{Uini}\ee

Integrating systems (\ref{Uqi}) with the initial conditions (\ref{Uini})
we can determine unknown vector functions $U^\mu_i(\tau)$ for $\tau>\tau(\lm_i)$
with the help of the functions $\Psi^{\pr\mu}_{i\pm}$ known from the initial
data on the segments (\ref{pseg}). This procedure is limited in $\tau$
by the ordinates of the points in which the trajectories $\s=\s_i(\tau)$
cross the characteristic lines $\tau\pm\s={}$const (Fig.\,2).
However we can continue this procedure for all $\tau$ if for every value of
$\tau$ (after calculating $U_i^{\pr\mu}$ from Eq.~(\ref{Uqi})) we determine
the functions $\Psi^{\pr\mu}_{i\pm}$ outside segments (\ref{pseg}) with
the help of Eq.~(\ref{qi}).

For this purpose we express functions $\Psi^{\pr\mu}_{(i-1)+}$ and
$\Psi^{\pr\mu}_{i -}$ from Eqs.~(\ref{con}) and (\ref{qi})
\be\begin{array}{c}
\Psi^{\pr\mu}_{(i-1)+}(\tau+\s_i)=
\Psi^{\pr\mu}_{i+}(\tau+\s_i)-
m_i\big[\gamma(1+\s'_i)\big]^{-1}U^{\pr\mu}_i(\tau),\\
\Psi^{\pr\mu}_{i-}(\tau-\s_i)=
\Psi^{\pr\mu}_{(i-1)-}(\tau-\s_i)-
m_i\big[\gamma(1-\s'_i)\big]^{-1}U^{\pr\mu}_i(\tau),\rule[3mm]{0mm}{1mm}
\end{array}\;\,i=1,2.\label{psipr}\ee
The similar relations for $i=3$ are
$$\begin{array}{c}
\Psi^{\pr\mu}_{2+}(\tau+\s_3)=(1+\s'_3)^{-1}\big[\Psi^{\pr\mu}_{0+}(\tau)
-m_3\gamma^{-1}U^{\pr\mu}_3(\tau)\big],\\
\Psi^{\pr\mu}_{0-}(\tau)=(1-\s'_3)\,\Psi^{\pr\mu}_{2-}(\tau-\s_3)-
m_3\gamma^{-1}U^{\pr\mu}_3(\tau).\rule[3mm]{0mm}{1mm}\end{array}$$

For solving the system (\ref{Uqi}) we are to determine the functions
$\s_i(\tau)$ for $\tau>\tau(\lm_i)$. These functions may be obtained
by taking some of the equalities (\ref{ch})
$$(1-\s'_i)\big(U_i(\tau),\Psi'_{(i-1)-}(-)\big)
=(1+\s'_i)\big(U_i(\tau),\Psi'_{i+}(+)\big)$$
and expressing $\s'_i$ in the following way (separately for $i=1,2$ and
$i=3$):
\be\!\!\frac{d\s_i}{d\tau}=
\frac{\big(U_i,[\Psi'_{(i-1)-}(-)-\Psi'_{i+}(+)]\big)}
{\big(U_i,[\Psi'_{(i-1)-}(-)+\Psi'_{i+}(+)]\big)},\quad
\frac{d\s_3}{d\tau}=1-\frac{\big(U_3,\Psi'_{0+}(\tau)\big)}
{\big(U_3,\Psi'_{2-}(\tau-\s_3)\big)}.
\label{sigma}\ee

The equations and systems (\ref{Uqi})\,--\,(\ref{sigma}) allow us
to continue the functions $\Psi^\mu_{i\pm}$ determined from the initial
conditions through Eqs.~(\ref{psi}) on the segments (\ref{pseg}).
The constants of integration in $\Psi^{\pr\mu}_{i\pm}$ are fixed from Eq.~(\ref{X=r}). This algorithm of calculating $\Psi^\mu_{i\pm}$
(with taking into account Eq.(\ref{sol})) solves the considered
initial-boundary value problem with arbitrary initial conditions
$\rho^\mu(\lm)$, $v^\mu(\lm)$.

\section{Stability of the rotational motions}

For the model ``triangle" the exact solutions of the equations of motion
(\ref{ort}), (\ref{eq}), (\ref{qqq}) describing the flat uniform rotation
of the system (with the quarks connected by hypocycloidal segments of
the strings) are known \cite{Tr,PRTr}.
World surfaces with the parametric representation
$\s_i(\tau)=\s_i={}$const and the form of closure condition (\ref{cl})
$\tau^*-\tau=T={}$const may be represented as follows
($X^1\equiv x$, $X^2\equiv y$):
\be X^0=\tau-\frac T{\s_3-\s_0}\s,\qquad X^1+iX^2=u(\s)\,e^{i\om\tau}.
\label{hyp}\ee
Here $u(\s)=A_i\cos\om\s+B_i\sin\om\s$, $\s\in[\s_i,\s_{i+1}]$;
$\om$ is the angular frequency of this rotation. Real
($\s_i$, $T$, $\om$, $m_i/\gamma$) and complex ($A_i$, $B_i$) constants
are connected by certain relations \cite{Tr,PRTr,ClTr}. Topologically different
types of solutions (\ref{hyp}) are characterized, in particular, by the integer
parameters $n$ and $k$, defined by the relations (if there are no singularities $\dot X^2=0$ in the segment $[\s_0,\s_1]$):
$$n=\lim\limits_{m_i\to0}\frac{{\s_3-\s_0}}{\s_1-\s_0},\qquad
k=\lim_{m_i\to0}\frac T{\s_1-\s_0}.$$
These values are equal to $n=3$, $|k|=1$ for the simple states in which
three quarks are connected by smooth segments of hypocycloids.
As we have said in the introduction the question about stability of such
motions remains open.

The above described method gives possibility to obtain (numerically, in general)
the world surface (\ref{hyp}) as the solution of certain initial-boundary value problem and also to research the stability of these motions solving
the problem with disturbed initial conditions $\rho^\mu(\lm)$ and $v^\mu(\lm)$.
As the simplest variations of initial conditions for
the model ``triangle" we consider an initial position of the string
in the form of a rectilinear triangle with the base $a$ and the altitude $h$
in the $xy$ plain. Initial velocities $v^\mu(\lm)$ correspond
to the uniform rotation of the system with the angular frequency
$\vec\om=\{0;0;\om\}$ about the center with coordinates $\vec\rho_c$
($\vec\rho$, $\vec v$ are notations for three-dimensional vectors in $R^3$,
the time coordinate in Minkowski space is $\rho^0(\lm)=0$, $v^0(\lm)=1$):
\be
\begin{array}{c}
\rho^1(\lm)=\left\{\begin{array}{l} a\lm,\\ a-b(\lm-1),\\ (a-b)(3-\lm);
\end{array}\right.\quad
\rho^2(\lm)=\left\{\begin{array}{ll} 0,& 0\le\lm\le1,\\
h(\lm-1),& 1\le\lm\le2,\\ h(3-\lm),& 2\le\lm\le3;\end{array}\right.\\
\rho^3(\lm)=0;\qquad \vec v(\lm)=\big[\vec\om\times(\vec\rho-\vec\rho_c)\big].
\rule[3mm]{0mm}{1mm}\end{array}
\label{ini}\ee
The quarks are placed at the corners of the triangle $\lm_0=0$, $\lm_1=1,\dots$

The systems of equations (\ref{Uqi}), (\ref{psipr}), (\ref{sigma}) are
analytically solvable only in some exceptional cases. Arbitrary initial conditions, in particular, the conditions (\ref{ini}) require numerical methods.
The numerical procedure of solving the initial-boundary value problem
for the relativistic string with massive ends is described in the paper
\cite{JVM}, the model q-q-q (technically more difficult) is considered
in Ref.~\cite{lin}.

In the present work we mention only the main points of our calculations.
At first we determine functions $\Psi_{i\pm}^{\pr\mu}$ on the segments
(\ref{pseg}) by formulas (\ref{psi}). Then we integrate the systems (\ref{Uqi})
simultaneously for all $i$ with the initial conditions (\ref{Uini})
and continue the functions $\Psi_{i\pm}^\mu$ with the help of expressions
(\ref{psipr}).

The results of computing are represented as projections of the world surfase
level lines $t=X^0(\tau,\s)={}$const at the $xy$ plain. This ``photos"
of sequential positions of the system are made with the step in time
$\Delta t$. For pictorial simplicity we consider only flat motions.

With an illustrative view the solution of the initial-boundary
value problem for the model ``triangle" is represented in Fig.\,3.
Here the parameters are: $m_1=m_2=m_3=1$, $\gamma=1$, the initial
position of the string has the form of the equilateral triangle
with $a=1$ and with zero initial speed of the string points,
that is with initial conditions (\ref{ini}) where
$a=2b=2h/\sqrt3=1$, $\vec v=\vec\om=0$.

In Figs.\,3a\,--\,3d the positions of the system are numbered in order of
increasing $t$ with the step in time $\Delta t=0.125$. The position of
the 3-th quark is marked by a circle. Owing to the symmetry of the system the quarks move along the three straight lines while the motion of the string
is rather complicated.

Let us turn to the stability testing of the rotational motions (\ref{hyp}).
We consider the initial conditions (\ref {ini}) slightly disturbing
these motions.

The detailed analysis of the problem was made for various parameters
of the model and initial conditions. The typical example of the solution
of the initial-boundary value problem for the ``triangle" model with close
but different masses and initial conditions (\ref{ini}) is represented in
Fig.\,4. We take the following parameters:
\be\begin{array}{c}
m_1=1,\quad m_2=1.5,\quad m_3=1.2,\quad\gamma=1,\quad\Delta t=0.15,\\
a=2b=0.5,\quad h=0.33,\quad\vec\rho_c=\{a/2;h/3\},\quad\om=2.
\rule[3mm]{0mm}{1mm}
\end{array}.
\label{val}\ee

The origin of the coordinates in Fig.\,4 coincides with the rotational center
(with the position vector $\vec\rho_c$). The values $\rho^\mu(\lm)$ and
$v^\mu(\lm)$ (\ref{val}) are close to those which give the exact hypocycloidal
solution (\ref{hyp}) for this system (with the same $m_i$, $\gamma$ and
the angular velocity $\om=2$). The last solution describes the uniform rotation
of the system with the string shape that is x-marked in Fig.\,4a. The string
positions for the solutions (\ref{ini})\,--\,(\ref{val}) are represented
by continuous lines. The positions of the third
quark with $m_3=1.2$ are marked by circles and numbered as in Fig.\,3.

The further evolution of the system is represented in Figs.\,4b\,--\,4d
(some time interval between Figs.\,4c and 4d is omitted for the sake of
space economy).

We see that when the system rotates the distances between the quarks
and the configuration of the string segments change and fluctuate near the
values corresponding to the motion (\ref{hyp}) (x-marks in Fig.\,4a).

This and a lot of other numerical experiments (with various values
$m_i$, the energy, $\rho^\mu(\lm)$, $v^\mu(\lm)$ and various types of
disturbances) show that the simple rotational motions (\ref{hyp}) with $n=3$,
$|k|=1$ are stable. That is small perturbations don't grow in time\footnote{
Testing the stability we, naturally, differ this problem from a uniform
motion of the system c.m. that takes place if the center of masses and
the rotational center in conditions (\ref{ini}) don't coincide.}.

We are to emphasize that the simple motions (\ref{hyp}) are steady with respect
to transforming into the ``quark-diquark" states with $n=2$, $k=0$ \cite {PRTr}
and merging two quarks into the diquark. One may obtain this transformation
only through rather strong disturbances\footnote{Note that large disturbances,
naturally, result in large changes of the motion. This fact is illustrated
by the example in Fig.\,3.}, for example, by reducing one of the sides of
the triangle (\ref{ini}). In particular, for the system with $m_1=m_2=m_3$
and initial quark speeds $v=0.5$ the motion remains ``simple" (similar to that
in Fig.\,4) if the relation $h/a$ of the triangle (\ref{ini}) with $b=a/2$
does not exceed the rather large critical value $(h/a)_{cr}\simeq2.93$.

In Fig.\,5 one can see the evolution of the described system ($m_i$, $\gamma$,
$\vec\rho_c$, $\om$ correspond to the initial quark speeds $v=0.5$) for
the initial relation $h/a=2.5$. The 3-rd quark is o-marked as previously
and the 2-nd quark is marked by the red square. The interquark distances
vary in an alternating manner.

With growing the energy of the system the value $(h/a)_{cr}$ grows.
If $h/a>(h/a)_{cr}$ the triangle orientation quasiperiodically changes
(the initial counterclockwise bypass direction of the numerated quarks isn't
preserved). In this motion the two nearest quarks revolve with respect to
each other but don't merge.

The stability picture of the ``triangle" baryon model radically differs
from that for the linear configuration q-q-q where any small disturbance
results in centrifugal moving away of the middle quark to one of the ends
(but also without merging and with quasiperiodical return of the middle
quark) \cite{lin}.

It is interesting to compare these two models in the case when one of the
masses $m_i$ is larger than the sum of two others. In this case the shape
of the triangle configuration for the simple rotational motion (\ref{hyp})
tends to a rectilinear segment with the position of the largest mass
at the rotational center (so the $\Delta$ configuration practically
tends to the q-q-q one) if the energy of the system decreases \cite{PRTr,ClTr}.

In Fig.\,6 such two motions of these two models are considered and compared.
For both models the quark masses are $m_1=m_3=1$, $m_2=3$ the tension is
$\gamma=1$ in the ``triangle" and $\gamma=2$ in the q-q-q configuration.
The initial shape
of the string is the rectilinear segment (for the $\Delta$ configuration it is
a particular case of the triangle (\ref{ini}) with $h=0$, $a=1/3$) rotating
so that the initial velocities of the 1-st and 3-rd quarks $v=0.5$ are connected
with $a$ by the relation $a=2(m_1/\gamma)v^2/(1-v^2)$ \cite{Ko,lin}.
The position of the 2-nd quark is slightly displaced with respect to the
center of rotation. In Figs.~5 and 6 $\Delta t=0.15$.

One can see that in the ``triangle" model (Figs.\,6a\,--\,6c) the
second (middle) heavy quark moves in the vicinity of the rotational center.
In the q-q-q model (Figs.\,6d\,--\,6f) the middle heavy quark moves away,
then it plays a role of rotational center for the string segment
between the two quarks and returns to the center of the system .
The latter case demonstrates that the rotational motions of the q-q-q configuration
with the middle quark at rest are unstable
(unlike ones in the ``triangle" model).

Note that the rotational motions (\ref{hyp}) include a set of so called
exotic states \cite{PRTr,ClTr} which contain points moving at the speed
of light. Such points (cusps) with singularities of the metric
$\dot X^2=X'{}^2=0$ are the typical phenomenon for the relativistic string
dynamics \cite{KlimN}. The stability analysis of these motions needs
other methods.

\section*{Conclusion}

In this paper the initial-boundary value problem for the string baryon
model ``triangle" is solved in general for arbitrary initial conditions.
This approach let us find that the simple rotational motions (\ref{hyp})
\cite{PRTr,ClTr} of this configuration are stable for all values of
the quark masses $m_i$ and the energy of the system.

Such a behavior radically differs from that for the linear string baryon
configuration q-q-q where rotational motions are unstable \cite{lin}.

Note that the initial-boundary value problem ``three-string" baryon
configuration is not solved yet. But the considered method after some
development may be used for this model too.

The results of our analysis justify that the simple rotational states
of the string model ``triangle" is applicable to describing the Regge
trajectories \cite{4B,PRTr} and give some additional arguments in favour of
this model\footnote{There are the arguments from the baryon Wilson loop point of view
in Ref.~\cite{Corn}.} in comparison with other string baryon models (Fig.\,1).

\end{document}